\documentclass[preprint2]{emulateapj}
\usepackage{graphicx}

\shorttitle{Bar imprints on M33 kinematics}
\shortauthors{Corbelli and Walterbos}

\begin{document}

\title{Bar imprints on the inner gas kinematics of M33}


\author{Edvige Corbelli}
\affil{INAF-Osservatorio Astrofisico di Arcetri, Largo E.Fermi, 5, 50125
Firenze, ITALY}
\email{edvige@arcetri.astro.it}

\and

\author{Rene' A. M. Walterbos}
\affil{Department of Astronomy, New Mexico State University, P.O. 
Box 30001, MSC 4500, Las Cruces, NM 88003}
\email{rwalterb@nmsu.edu}

\begin{abstract}

We present measurements of the stellar and gaseous velocities in the
central 5$'$ of the Local Group spiral M33. The data were obtained
with the ARC 3.5m telescope. Blue and red spectra with resolutions
from 2 to 4\AA\ covering the principal gaseous emission and stellar
absorption lines were obtained along the major and minor axes and six
other position angles. The observed radial velocities of the ionized
gas along the photometric major axis of M33 remain flat 
at ~22 km~s$^{-1}$
all the way into the center, while the stellar velocities show a
gradual rise from zero to 22 km~s$^{-1}$ over that same region.  The
central star cluster is at or very close to the dynamical center, with
a velocity that is in accordance with M33's systemic velocity to
within our uncertainties.  Velocities on the minor axis are non-zero
out to about 1$'$ from the center in both the stars and gas. Together
with the major axis velocities, they point at significant deviations
from circular rotation. The most likely explanation
for the bulk of the velocity patterns are streaming motions along a
weak inner bar with a PA close to that of the minor axis, as suggested
by previously published IR photometric images. The presence of bar imprints
in M33 implies that all major Local Group galaxies are barred. The
non-circular motions over the inner 200 pc make it difficult to
constrain the shape of M33's inner dark matter halo profile. If the
non-circular motions we find in this nearby Sc galaxy are present in
other more distant late-type galaxies, they might be difficult to
recognize.

\end{abstract}


\section{Introduction}

The central regions of late type spiral galaxies are still quite
puzzling places. Often they show a young, massive, compact star
cluster and no clear evidence of a stellar bulge or bar
\citep[e.g.][]{bok02} leaving open the question of a fueling 
mechanism for star formation in their shallow central potential.  
The central region of the Local Group galaxy M33 is the posterchild 
of such an environment. It hosts a bright blue
compact cluster at its center but, unlike other bright 
Local Group members, no definitive evidence 
of a bar or bulge has been found yet. Being
nearby \citep[840~kpc,$1''$=4~pc,][]{fre01} 
and at favorable inclination \cite[52$^\circ$,][]{cor00}, M33
allows however a detailed look at the spatial distributions and 
kinematics of various components in the central region.

The mass of the blue compact nuclear cluster in M33 has been estimated
to be of order $10^6$~M$_\odot$ \citep{kor93,lau98}. Its small
mass-to-light ratio is indicative of a young population of stars (age
$\sim 10^7$-$10^8$~yr) concentrated in the compact core
\citep{kor93}. An age spread is likely to be present across the
cluster, due to a possible sequence of accretion events. However,
there are models which predict a single-age stellar population by taking
into account the role of dust in processing the UV/optical stellar 
spectrum \citep{gor99}.  From the theoretical point of view
it is unclear whether compact nuclear clusters are born and manage to
survive at the dynamic center or they develop non-steady
motions around the mass centroid or migrate towards the center and
eventually get disrupted as they approach it
\citep{mil92,por03,mil04}. In the latter cases the dynamic center
would not coincide with the position of the bright cluster, but given
the shallow potential of M33 \citep{vdm06} it is unclear that this 
has happened for this galaxy.  There has been no accurate and
conclusive spectroscopic determination of the dynamical center via
stellar absorption lines or emission line gas mapping of the central
1~kpc region of M33. Rubin \& Ford 1980 noticed a possible small
displacement of the dynamic symmetry center with respect to the
location of the compact nuclear cluster.  Locating the dynamical
center of M33 has been one of the goals of our project even though we
shall discuss in this paper the difficulties of reaching a definite
conclusion on this issue.  With the nuclear cluster light taken out,
the photometric center of M33 is difficult to identify because of the
presence of dust.  M33 lacks a supermassive black hole at the center,
the upper limit to the mass being as low as 1500~M$_\odot$
\citep{geb01,mer01}.  Given the shallow potential well of the center
of M33 this is not a surprise although the downward extrapolation of the
relationship between stellar velocity dispersion and supermassive
black hole mass predicts a black hole mass for the cluster that is
well above the quoted limit.

Understanding the innermost region of M33 is important also for
possible constraints to cosmological models of structure formation, in
particular to distinguish cuspy dark matter halo models from
isothermal core models.  In a recent determination of the rotation
curve from the CO J=1-0 line data at $R>0.5$~kpc, Corbelli (2003)
discusses cosmological models compatible with the data and finds that
an additional mass component is required in the innermost 1~kpc. What
is the nature of this component? Can measurements of the kinematics of
stars in the innermost 1~kpc region help in solving some of these issues?

Photometric evidence for a bulge  has been presented for M33
but this is far from being conclusive.  Exponential profiles
typically provide a good fit to the stellar surface brightness in the
large-scale disk but in the inner few hundred parsecs the brightness
exhibits an excess with respect to the inward extrapolation of the
disk exponential law \citep[e.g.][]{bot92,reg94}. It is unclear
whether this is due to an excess of stellar density distributed in a
disk or to a bulge/spheroidal light distribution.  In the latter case
Regan \& Vogel (1994) derive a large angular size spheroid (8$'$ or
2~kpc effective radius) from photometric fits to the disk and
spheroidal light distribution.  Minniti et al. (1993) have claimed a
photometric detection of a smaller central component, a bulge with an
effective radius of 0.5~kpc, through JHK-band imaging of the central
region of M33.  This is perhaps a pseudobulge since it underwent a
star formation episode less than 1~Gyr ago. They find a displacement
of the center of symmetry of the bulge with respect to the nuclear
cluster by about 20$''$ to the southeast.  They mention dust as a
possible origin of this photometric center shift but we note that the
large OB stellar association south-east of the nucleus might also
cause some displacement of the light center. Recent GALEX far-UV and
near-UV mosaics of M33 \citep{thi05} have also shown an excess of
light in the innermost 0.5~kpc region, in agreement with optical
photometric results and with the high level of star formation detected
by infrared and H$\alpha$ surveys (see Magrini
et al. 2007 and references therein).

A change in the ellipticity of the isophotes between the inner
and the outer regions of a disk galaxy is ofter used to infer
the presence of weak bars in nearby galaxies and to measure the
bar strength even in more distant systems \citep{abr99}.
Larger IR maps of M33 \citep{reg94,blo04} point at a weak bar which becomes
visible after subtraction of the exponential disk and a spheroidal
light distribution. The bar would have a position angle close to
90$^\circ$ and projected radial extend smaller than 6'.  However an
alternative explanation to the bar-like emission is a possible change
of the pitch angle of the the spiral arms at smaller radii.  Fourier
decomposition of galaxy images in the blue has also revealed a bar
structure \citep{elm92}.  The metallicity, which has a negligible
gradient in the entire star forming disk of M33 \citep{cro06,mag07},
also rises suddenly in the innermost 0.5~kpc indicating a possibly
complex star formation history.

It is remarkable that in spite of this large number of photometric
studies of the innermost region of M33, there is not a similarly
detailed study of the stellar and gas kinematics. Colin \&
Athanassoula (1981) noticed an appreciable velocity asymmetry in the
inner $5'$ velocity field using the ionized gas as a kinematic
tracer. This however has not been confirmed by the CO rotation curve
\citep{wil89,cor03}. In a conference paper, Rubin \& Ford (1980)
presented emission-line gas spectroscopy at $R<1'$ along the major axis
only. They find a constant radial velocity all the way into the
center, an intriguing result that has not received further follow-up
or interpretation.  The published HI data \citep{new80,deu87} do not
have sufficient angular resolution to assess the kinematics at small
distances from the center and within 200~pc radius
only a few molecular clouds have been detected \citep{wil90,hey04},
insufficient to measure a rotation curve. Therefore, the central gas
kinematics has to be investigated via line emission from HII regions
and diffuse ionized gas, which fortunately is very abundant
(see Figure 1).  Stellar velocities, to our knowledge, have
never been measured. The intrinsic difficulties of accurate
kinematical measurements in the central region of this nearby galaxy
compared to e.g. M31, lie in the relatively fainter stellar light
(except at the location of the nuclear cluster) and in the slow rising
of its rotation curve which implies velocities below 30~km~s$^{-1}$ in
the inner ~0.5 kpc region. These are of the same order or lower than
the expected stellar velocity dispersions.

In this paper we will discuss the results of a set of observations
which outline the complex kinematics of the central region of M33.
These will hopefully complement other information available for this
nearby galaxy, to further establish it as a template for interpreting
observations of less resolved, more distant late type galaxies.  In
Section 2 we outline our observations of the stellar and gaseous
kinematics in the inner 0.5~kpc of M33. Results are presented in
Section 3, while Section 4 discusses possible dynamical mass models 
to explain the data. We conclude with a summary of our most relevant 
results in Section 5.

\section{Observations and data reduction}

We obtained spectroscopic observations of the central region of M33
with the ARC 3.5-m telescope in various short observing runs in 2002,
2004, 2005, and 2006 using the DIS long-slit spectrograph. The two
cameras in DIS covered wavelengths from 3700-5500\AA\ (2004) or
4300-5500\AA (other years) in the blue, and 5600-7200\AA\ in the
red. The dichroic which splits the light in blue and red components is
around 5400\AA\ which implies low sensitivity in the region
5300-5500\AA. Spectra of the Ca triplet near 8500\AA\ were obtained
for the major axis (PA=22$^\circ$) in 2005. The slit width for all
spectra was 1.5$''$.  The slit length was 5$'$ which corresponds to 1.2 kpc
at a distance of 840~kpc. Between the 2002 and 2004 observations
the optics of the spectrograph were replaced resulting in changes to
the spatial scale from 0.54$''$/pixel to 0.4$''$/pixel. The 2006
observations were taken with new high resolution gratings in
place. The spectral resolution of the various spectra (FWHM) varied
from 2.0 to 4.0 Angstrom. Seeing was typically between 1$''$ and
2$''$. Spectra were obtained at position angles 18$^\circ$, 22$^\circ$
(nominal major axis), 112$^\circ$ (minor axis), 96$^\circ$ (suspected
bar major axis), 59$^\circ$, 130$^\circ$, 146$^\circ$, and
160$^\circ$. The PA=18$^\circ$ data were combined with PA=22$^\circ$. The
slits are shown overlayed on the H$\alpha$ emission image in
Figure 1. In some cases we obtained two spectra shifted in position to
extend the spatial coverage to more than 5$'$ along a given PA. Major
and minor axis spectra were observed a few times with somewhat
different centering for the nucleus. In addition, several spectra
parallel to, but offset from the major axis spectrum by 1$''$ to 8$''$, and
one at 142$''$ SE offset were obtained. These are not shown in Figure 1
to avoid clutter. The net exposure times ranged from 20 to 90 minutes,
split in double or triple exposures to eliminate cosmic rays in
post-processing. The shorter exposure times apply to the intermediate
position angles, while the longer times apply to separate major
and minor axis spectra. For all exposures the galaxy was accurately
centered in the slit using the slitviewer camera before offsets were
done as needed.

Several stars of spectral type F5V through K3III were observed as
templates for the stellar velocity measurements.  The blue spectra of M33
were predominantly used to measure the stellar velocities, although
the H$\beta$ and [OIII] emission lines did provide a check on the
H$\alpha$, [NII], and [SII] velocities from the red spectra. In the
2006 run of PA= 130$^\circ$, 146$^\circ$, and 160$^\circ$ the red
camera was not available and we include the H$\beta$ emission line
velocities for those cuts. The Ca triplet observations provided a more
accurate zero point for the stellar velocities than the blue spectra
due to the presence of numerous bright sky lines in the IR. Since we
wanted to measure the red emission lines, and because the CCD in use
for our observations showed significant fringing in the IR and had
rather low sensitivity around 8500\AA, we only observed the major axis
at this wavelength range. These data did enable us to anchor the
velocity of the cluster and bright disk part on the major axis, and to
measure stellar velocity dispersions along the brightest part of the
major axis.

The large angular size of M33 implies that separate sky exposures had
to be obtained to subtract sky lines. We observed blank sky regions
each night between M33 exposures. HeNeAr lamp exposures were obtained
regularly for wavelength calibration. For the red spectra, velocity
zero points were determined from the 6923\AA\ sky line for the red
emission lines, and from 4 sky lines near the Ca triplet. The blue
region of the spectrum has no bright sky lines near the region of
interest and the blue spectra's zero points were set by aligning them
with the corresponding red spectra. We estimate the final uncertainty
in the zero points to be $\pm 4$~km~s$^{-1}$. 

All data were reduced using standard IRAF processing techniques.
Elimination of cosmic rays was done using the crreject option in
IRAF. This works well for most of the spectra, but can introduce
spurious effects for the bright nuclear star cluster, due to slight
changes in seeing, the effects of refraction in the earth's
atmosphere, or in exact centering of the nucleus in the
slit. Therefore, measurement of the gas velocities close to the
nucleus (shown in Figures 8 and 9) and of the Ca triplet data for the
star cluster was done on single spectra. Fringing in the Ca triplet
data was reduced satisfactorily by flatfielding the spectra with the
quartz flat.

In addition to the spectroscopic observations we also have a deep
H$\alpha$ image of M33 obtained with the Burrell Schmidt (Hoopes \&
Walterbos 2000) and we obtained short-exposure, sub-arcsec seeing
H$\alpha$, [OIII], and continuum images of the central 4.6$'$ with the
SPICAM imaging camera on the ARC 3.5-m telescope. The latter images
provide useful information on the morphology and location of dust
lanes, OB assocations, HII regions, and other emission-line
sources. Some of the images are shown in Figure~1.

From 21-cm data the inferred systemic velocity of M33 is
-180~km~s$^{-1}$, the major axis orientation is PA=22$^\circ$, and the
disk inclination is $i=52^\circ$ \citep{cor00} and we shall use these
values throughout the paper.  The approaching half of M33 is west of
the major axis.  In showing the velocities along a slit position, our
convention is that negative radial distances refer to the east side of
the galaxy and increase towards the west side.

\subsection {Gas velocity and velocity dispersion measurements}

The velocities of the ionized gas were determined on blue and red
spectra, measuring the H$\beta$, [OIII]5007, H$\alpha$, [NII]6783,
[SII]6716, and [SII]6731 lines. In the red, the four lines were
measured simultaneously to determine one redshift, and individually to
test internal consistency. Close to the nucleus [NII]6583 remained the
only line visible due to a strengthening of the [NII]/H$\alpha$ ratio
and the strong stellar H$\alpha$ absorption of the stellar
nucleus. Outside the nucleus H$\alpha$ was the strongest
line. Velocities from the red spectra typically have better accuracy,
although the H$\beta$ velocities generally agreed well with the
H$\alpha$ results. Velocities from the [OIII]5007 line line did not
always agree with the Balmer line measurements, likely indicating real
differences. The [OIII] emission is enhanced in shocks and high
excitation regions and it is possible that the [OIII] velocities
preferentially probe a less quiescent gas phase; this will be
addressed in a second paper where we look at the excitation ratios of
the gas in the central region. All the lines were measured with the
SPLOT routine in IRAF.  From consistency checks between different
exposures of the same P.A., overlapping sections between spectra, and
internal consistency between the 4 emission lines in each red spectrum
we infer that typical uncertainties in the emission line velocities
are of order 4~km~s$^{-1}$ (including zero point uncertainties). 

\subsection{Stellar velocity and velocity dispersion measurements}

In order to derive the M33 stellar velocities along the line of sight
we used a cross-correlation technique \citep[as described by][]{kur98}
and implemented in the RVSAO set of programs in the IRAF
environment. This technique is based on the original power spectrum
method developed by Tonry \& Davis (1979) and is suitable for rotation
curve measurements when line profiles are Gaussian and no complex
velocity structures or multiple components are present \citep{bar00}.
The faint and low mass stellar disk of M33 does not show evident signs
of complexity and furthermore the high spatial resolution achieved in
this nearby galaxy minimizes the possibility of multiple components
overlapping along the line of sight. Our spectral resolution is such
that no precise information on the stellar velocity dispersion (of
order 25-30~km s$^{-1}$, e.g. Kormendy \& McClure 1993) could be derived
from the blue data, but the Ca triplet data did provide some results.
The stellar disk kinematics were obtained from absorption features
present in the highest resolution spectra across the wavelength range
running from about 4400\AA\ to 5500\AA.  In order to investigate the
reliability of $H\beta$ absorption line, often mixed with emission
line gas, we have also used the cross-correlation technique in the
restricted wavelength range 5030\AA\ to 5500\AA\ and find consistent
results: we shall refer to this restricted wavelength range
correlation as the `Mg side' correlation. All velocities were
corrected to V$_{helio}$ using the program rvcorrect in IRAF.

A zero point offset in the velocity scale arises in the resulting
velocities because of the varying flexure of the spectrograph as it
rotates at the Nasmyth focus. The stars observed for cross-correlation
are never observed at the same rotator angle of M33 implying a varying
relative zero point after wavelength calibration is applied. We
aligned the velocity curves setting the cluster velocity to
-180~km~s$^{-1}$, as suggested by Ca triplet line analysis.

A number of stellar templates were observed for each instrumental
set-up. Spectra were rebinned along the spatial direction to obtain a
nearly constant signal-to-noise ratio along the slit. The
uncertainties shown in the results are those provided by the XCSAO
routine outputs and they do not take into account possible template
mismatch. Along the disk, different templates gave consistent results
to within a zero point shift only. The stellar nuclear cluster is
bluer than the disk, likely different in age, and it may host a
mixture of young and old stellar populations. Some template mismatch
between nuclear cluster and disk then can cause higher uncertainties
in the velocity zero point of the cluster compared to that of the disk
when blue spectra are used. This results in an increased scatter
eventhough the measured radial dependence of the stellar velocities in
M33 was generally consistent along the disk. This is evident for
example around the nuclear cluster region and along the minor axis
(shown in Figure~3, top panel).  The Ca triplet lines are narrow and
strong and vary only slowly with metallicity and star
temperature. They can therefore be used even if a mixture of stellar
populations is present \citep{kor93}. Outside of the nucleus however,
as the lines become weak, sky line contamination and remaining
fringing become a problem. We therefore obtained results only for the
cluster and central disk region on the major axis.  To best estimate
the radial velocity of the star cluster we measured the position of
sky lines and the wavelengths of the Ca triplet lines in the
single-exposure wavelength calibrated spectrum with the best
seeing. This was done through a fit to several sky lines and to the
sharp troughs in the triplet absorption lines. This way we avoid any
influence of the stellar template that is used in the
cross-correlation method on the derived radial velocity. In addition,
by tying the absorption velocity to the sky line wavelengths we
eliminate effects of a zero point offset in the velocity scale
associated with varying flexure in the spectrograph as it rotates at
the Nasmyth focus. The resulting velocity we derived for the cluster
was indistinguishable within the uncertainty from the systemic
velocity of -180 km~s$^{-1}$. We estimate our zero point uncertainty
to be about $\pm4$ km~s$^{-1}$. Through cross-correlation with stellar
templates in the restricted wavelength range 8480\AA\ 8680\AA\ we also
measured stellar radial velocities close to and in the cluster region
(Figure 2).

\section{Kinematics of Gas and Stars in the central region}

The M33 data presented in this paper probe velocities at high spatial
resolution and very close in to the nucleus compared to typical
rotation curve measurements obtained from the ground for almost all
other galaxies.  A few issues that pertain specifically to the central
region of this nearby low luminosity disk galaxy need to be
underlined. First, the observed radial velocities and the velocity
dispersions are quite low. Radial velocities are less than 30 km~s$^{-1}$. 
This is of the
same order as might be expected for velocity gradients across spiral
arms and for velocities in expanding HII shells around regions of star
formation. We will focus here on systematic velocity patterns rather
than individual features. Second, the typical uncertainty in the zero
point determinations is about $\pm 4$ km~s$^{-1}$. This implies that
any apparent asymmetries of this order in the measured radial
velocities from one side of the center to the other may not be
real. Third, the scaleheights of the gaseous and stellar disks are
expected to be much larger than the closest distances to the central
region we can probe: 1$''$ corresponds to 4~pc. Therefore, projection
effects can lead to misinterpretation of velocities very close to the
center, because the gas could be further away from the center even if
seen projected close in. We don't know the scaleheights of the various
components in the mass modeling and this also becomes a bigger problem
close to the center where the nuclear cluster is a compact source in a
likely much thicker gas disk.

\subsection{Emission line velocities of the ionized gas}

The major and minor axis velocities of the ionized gas in the inner
region of M33 are shown in bottom panels of Figures 2 and 3. The most
remarkable feature of the major axis velocities is that they are
constant at $\sim 22$~km~s$^{-1}$ on average from $2'$ (0.5~kpc) down
to within $2''$ ($\simeq 8$~pc) from the center, in agreement with the
results by Rubin \& Ford (1980). The minor axis velocities are
characterized by significant non-zero velocities within 1-1.5$'$ from
the center, and reach maximum values similar to those observed along the
major axis.  Even though velocities have a regular pattern along the
minor axis, they are far from being constant. Here velocities are not
symmetric with respect to the blue cluster center which 
appears to be located at or close to the dynamical center (see
also Section 3.2 and 3.3). On the west side of the minor axis
velocities vary by about 50~km~s$^{-1}$ across the innermost $1'$
region.  The minor axis is close to the suspected bar direction
(estimated at approximately PA=96$^\circ$ from
Fig.1$(c)$ in Block et al.2004 or Fig.1$(b)$ in Elmegreen et al. 1992). 
We show the ionized gas velocities
for PA=96$^\circ$ in the middle panel of Figure 4. The velocities in this
direction are much like that of the minor axis, with
somewhat reduced amplitude. The velocities East of -1$'$ drop
below systemic as expected, given we see the
approaching side of the disk here.

The other two panels in Figure~4 show velocities along PA=59$^\circ$ and
PA=146$^\circ$, while Figure~5 repeats the PA=146$^\circ$ data
averaging all four red lines, and adds H$\beta$ measurements along
PA=146$^\circ$ and the neigboring directions PA=130$^\circ$ and
160$^\circ$. A few points are worth noticing about the radial
velocities at the various intermediate PAs. First, the radial velocity
maxima reach values as high as observed on the major axis. This
would not be expected if major axis velocities were due to gas in
circular motion around the center. Specifically, if the rotation curve
were flat inside 0.5 kpc as suggested by the major axis velocities, we
ought to see constant velocities with a reduced amplitude and a sign 
change at the center, for all the intermediate PAs. In the inner $\pm1'$ 
velocities are instead asymmetric with
respect to the center, especially along PA=146$^\circ$. Here east side
velocities rise to about 30~km~s$^{-1}$ and abruptly decrease around
-1$'$ while on the west side they stay close to systemic for D$<1'$.

In Figure~7 we show line of sight velocities observed along directions
parallel to major axis. The slits  cut the minor axis at
different projected distances from the nuclear cluster center.  
Major-axis velocities show the steepest jump when the minor axis
is crossed. The systematic shift in the velocity at the minor axis
crossing that is evident in the various cuts, is in agreement with the
non-systemic velocities observed on the minor axis.  We don't notice
any peculiarity in the velocity curve which samples regions at larger
galactocentric distances along a direction parallel to major axis
(142$''$ east offset). The velocities appear normal for disk rotation
and the intersection with the minor axis occurs at the systemic
velocity this far out.

We tried to measure the velocity dispersions implied by the widths of
the emission lines. The highest S/N exposures of the major and minor
axis gas velocities in the red were from 2002. These data have a
spectral resolution (FWHM) of about 2.5\AA, corresponding to
115~km~s$^{-1}$, or a 1 sigma dispersion of about 50 km~s$^{-1}$. We
measured the dispersions by deconvolving the widths of the 4 principal
red lines (H$\alpha$, [NII]6548, [SII]6716+6731) with the widths of
nearby sky lines in the same frame, or with the widths of lines in the
arc spectrum at exactly the same location of the data. Typically, 
in cases where the deconvolution of the line
widths gave a consistent result, the derived dispersions were 12 to
20~km~s$^{-1}$ with only very few locations of higher values. There
were also several cases where no discernable broadening could be
measured. The dispersion due to thermal motion of the gas, for
temperatures of 5000 to 10,000K, would be 6 to 8~km~s$^{-1}$ for the
Balmer lines.

\subsection{The stellar velocity pattern}

In the top panels of Figures~2 and 3 we show the stellar velocities
measured along the major and minor axis. For the major axis, we
include results from the Ca triplet and from the blue spectra, while
the minor axis has only velocities derived from the blue
spectra. The use of the whole wavelength range in the blue or only of
the Mg side (see Section 2) gave consistent results. Velocities
along the major axis rise much more slowly for the stars than for the
gas. On the minor axis, velocities are close to or at systemic,
although we do note that average velocities within $\pm 1'$ from
the center are positive with respect to the systemic velocity, a trend
which is also observed for the gas.  This assumes that the star
cluster is at the systemic velocity, a result derived from the Ca
triplet measurements. The increased scatter in the minor axis stellar
data compared to the major axis data is likely due to increased
uncertainty regarding the relative zero points of the nuclear star
cluster versus the disk velocities, as discussed in Section 2.

We also measured stellar velocities at PA=130$^\circ$, 146$^\circ$,
and 160$^\circ$. These data were obtained with a new blue high
resolution grating in place in 2006. The results are shown in the
panels of Figure~6. The radial velocity patterns for the stars along
these directions are quite similar to those of the gas (Figure 5) but
of lower amplitude.

The Ca triplet data also provided some information on the stellar
velocity dispersions. We find 20-23 km~s$^{-1}$ for the star cluster,
in good agreement with previous results \citep{kor93,geb01}, and
values in the range 28 to 35 km~s$^{-1}$ along the inner disk major
axis.

\subsection{Velocities in the innermost 60~pc region}

To better characterize the emission line velocities close to the
nucleus, we measured each of the four red emission line on individual
exposures of the major and minor axis. This was done to avoid any
smearing or loss of information that might be introduced close to and
across the bright nucleus in the combination of frames during the
cosmic ray elimination process. We measured the velocities every
single spatial pixel over a region of -15$''$ to +15$''$ along the
major and minor axes. The velocities derived this way were all shifted
in zero point using the 6923\AA\ sky line, and then averaged for each
pixel when all frames have consistent values i.e. an rms dispersion
around the mean less than 25~km~s$^{-1}$. Measurements are not
averaged and refer to single frames when the rms dispersion was larger
or lines were hardly measureable due to blending with stellar
absorption or to intrinsic weakness. [SII] velocities are an average
of the velocities derived for the 6716\AA\ and 6731\AA\ lines when
they give consistent values. In the opposite case individual
measurements of the two [SII] velocities are shown, rather than the
average and rms dispersion. Results are shown in Figures~8 and 9.

Data along the major axis show that velocities remain constant in
H$\alpha$ and [NII], while the [SII] velocities seem to increase near
the center. The latter result is uncertain given the large scatter in
individual measurements. The only emission line that remains visible
across the stellar nucleus is [NII]6854. Even this line is however of
limited S/N due to underlying very bright stellar continuum; the
[NII]6584 line sits just at the edge of the deep stellar H$\alpha$
absorption associated with the cluster. The [NII] velocities on the
major axis suggest that the systemic velocity is not measured at the
position of the cluster but about 1$''$ to the SW of it (along the
major axis). This offset was also noted by Rubin \& Ford (1980).  On
the minor axis we find the same value of the gas radial velocity we
measure at the cluster position on the major axis.  The results are in
agreement and point out to a gas velocity of order -200 km~s$^{-1}$
which could imply that the dynamical center is not exactly at the
cluster center. However, the presence of similar velocity jumps over
short distances on the cuts parallel to the major axis (Figure~7) and
the possibility of outflows and non-circular motions imply that it may
be difficult to pinpoint the exact dynamic center of M33.  The nucleus
might be "wandering" around the dynamic center, either falling in or
oscillating (e.g. Miller $\&$ Smith 1992).

Another somewhat striking aspect of Figure~9 is the negative velocity
feature on the minor axis at D$\simeq 8''$ in western
direction.  This is measured quite accurately in H$\alpha$, and
reproduced at the intersection with the minor axis on the cut 8$''$
offset from the major axis in Figure 7 (top panel). The feature is
about 30~km~s$^{-1}$ deep with respect to systemic extending over
$\pm3$ arcsec or $\pm$12~pc.  No expanding HII shell is visible at
this location. In our H$\alpha$ image there is a combination of
diffuse emission and faint filamentary structures in this region, much
like the emission over most of the area covered by our spectra, but
not a single large shell. It is possible that the feature is part of
the general oscillatory patterns we see in the gas motions here, that
may be associated with shocks or with the bar (see Section 4). A
more speculative option is that the dip is associated with a possible
keplerian fall off or rotation connected to a black hole at or near
that location. However, at our slit positions we did not notice any
evident increse in the stellar velocity dispersion. Any meaningful
discussion on this issue requires full 2-D velocity field mapping of
the central region with high spatial and velocity resolution.

\section{Mass modelling of the central region}

A comparison of the stellar and gas major axis velocities points to a
systematic difference: the stars seem to rotate more slowly than the gas
does.  This
behaviour suggests that asymmetric drift reduces the stellar rotation
curve, with the effect being most pronounced close to nucleus, as
observed.  There are two other noticeable features in the ensemble of
gas radial velocity plots shown in the previous Sections.  $(a)$ Going
radially outwards from the nuclear cluster there is not a particular
direction that has the maximum or minimum radial velocities.  The
major axis direction cannot be easily defined then: for each slit
position crossing the center the maximum observed radial velocity is
of order 20-30~km~s$^{-1}$ in the radial range $\pm 1'$.  Continuous
tilting of the orbits i.e. a non defined preferential direction for
the major axis is not sufficient to explain the observed pattern.
$(b)$ Velocities along a slit passing through the center do not show a
clear symmetry with respect to the nuclear star cluster except along
PA=22$^\circ$, the major axis of the large scale disk.  Asymmetries
with respect to the nuclear cluster might be present in the stellar
velocities as well, especially along PA=146$^\circ$, even though
uncertainties in the results of the cross-correlation tecnique and in
the central cluster velocity limit our ability to firmly establish
their amplitude.  Thus models which consider pure circular orbits 
might be inappropriate for the innermost region of M33. 
However, we will first explore these models, where circular components
make the largest contribution to the observed radial velocities, attributing
any gas residual velocity to local peculiar motions such as radial 
inflows/outflows, shocks, etc. In this
framework asymmetric drift should provide the likely explanation for
any difference between gas and stellar velocities and gas velocities
should be consistent with a dynamical mass model. 
In the second part of this Section we shall examine if
possible asymmetries in the stellar and gas velocity pattern can be
related to the displacement of a small bulge, as suggested by Minniti
et al.  (1993), which would be the dominant mass component in the
innermost region. At the end of this Section we interpret the emission
line gas kinematics inside $\pm 1'$ in terms of large non-circular
orbits around the dynamical center and discuss the possible role of a
weak stellar bar in determining the gas flow pattern and its
deviations from pure circular orbits.

\subsection{The axisymmetric potential}

In this Section we consider circular orbits for the gas and stars,
centered on the nuclear star cluster. The star distribution is
symmetric around the azimuthal direction and about the disk equator.
We shall assume that the orientation of the orbits along the line of
sight follows that of the large scale disk, namely they are inclined
by $i=52^\circ$ around the position angle of the major PA=22$^\circ$.
We assume at first that the circular rotational velocity is traced by
the gas emission lines observed along the nominal major axis,
corrected by the inclination of the disk along our line of sight.  For
$R<2'$, gas radial velocities imply a flat rotation curve with 
an amplitude of 29~km~s$^{-1}$.

In Figure 10 we show these velocities, averaging and binning red and
blue emission line velocities along along PA=22$^\circ$ (triangle
symbols). The open stars are stellar rotational velocities along the
same direction. In Figure 10 the CO J=1-0 rotation curve is also
plotted. This is obtained from azimuthal averages of data 
within $\pm 45^\circ$ of the major axis (Corbelli 2003).  The azimuthal
averages of the CO velocities are larger than optical line emission
velocities at radial distances around 0.5~kpc. This may be due to
peculiar motion (e.g. crossing of spiral arms) along various directions 
or to the fact that PA=22$^\circ$ may not
be exactly the major axis direction.  The
rotation curve is traced by optical emission line velocities:  is
flat inside $R=2'$ and the mass density should decline as $1/R^2$.  
The stellar circular
velocities follow the gas rotational velocities after 
asymmetric drift corrections are applied.
Considering the mean of 
the radial velocity times the velocity perpendicular to the plane as
constant with height above the plane, from the mass
volume density $\rho_s$ we can estimate the drift following equation 4.33 
of Binney $\&$ Tremaine (1987) as:

\begin{equation}
V^2_c-V^2_s=-{{\sigma_R^2}}\lbrack 1-\sigma^2_\phi/{{\sigma_R^2}}
+2\ \partial {\hbox{ln}}\rho_s/\partial {\hbox{ln}}R \rbrack 
\end{equation}

\noindent
where  $V_c$ and $V_s$ are the circular velocity and observed stellar 
rotational velocity respectively,  $\sigma_R$, $\sigma_\phi$, and 
$\sigma_z$ are the radial, azimuthal and vertical velocity dispersion. 
In the above equation we have assumed that 
$\partial {\hbox{ln}}(\sigma_R^2\rho_s)/\partial {\hbox{ln}}R \simeq 
2 \partial {\hbox{ln}}\rho_s/\partial {\hbox{ln}}R $.
Using also the assumption that $\sigma_\phi^2\simeq \sigma_z^2 \simeq 
0.45{{\sigma_R^2}}$ (Binney $\&$ Tremaine 1987, eq.3.76), we have:

\begin{equation}
\sigma_z^2={V^2_c-V^2_s\over 2.2 \lbrack -0.65
-2\ \partial {\hbox{ln}}\rho_s/\partial {\hbox{ln}}R \rbrack}
\end{equation}
 
The observed stellar rotation is well fitted by $V^2_s\simeq 200
R^2$, where $V$ is in km~s$^{-1}$ and $R$ in arcmin. Assuming that
the observed gas velocities trace $V_c$ and hence the gravitational
potential, we can infer the mass density $\rho_s$ from the gas kinematics.
The flatness of $V_c$ in the innermost region of M33 and the above
equation implies $\sigma_z$ of order 10~km~s$^{-1}$. This is too a
small value given the observed stellar dispersions in the disk
(of order 30~km~s$^{-1}$), away from the nuclear cluster. Therefore
the mismatch between a flat rotation curve for the gas and a rising
rotation curve for the stars cannot be only the result of asymmetric
drift. Likely the flatness of the radial velocities distribution along
PA=22$^\circ$ is not due to pure circular motion but to a
combination of rotation and some streaming motions in a
non-axisymmetric potential. This picture also provides a natural
explanation for the non zero velocities observed along the putative
minor axis at PA=112$^\circ$ and we shall discuss it in Section 4.3.
A slow rising gas rotation curve reaches a better agreement with the
stellar rotation curve affected by asymmetric drift.  Since stellar
velocity dispersions in the disk are 30 km~s$^{-1}$ or higher, it is
feasible to assume $\sigma_z> 20$~km~s$^{-1}$. Since $V^2_c-V^2_s$
is of order 29~km~s$^{-1}$ close to the center and decreases further
out, then eq.(1) implies that the derivative of the stellar
density should be:

\begin{equation}
\partial {\hbox{ln}}\rho_s/\partial {\hbox{ln}}R > -0.8
\end{equation}

This condition implies that asymmetric drift can explain the differences
between the stellar and the gas rotation curve when the last one is 
rising. This is inconsistent with the hypothesis that
optical gas velocities observed along PA=22$^\circ$ effectively
trace the rotation curve in an axisymmetric potential. An almost flat
rotation curve requires in fact the above derivative to be of order $-2$.
There are other difficulties related to the interpretation that the
optical gas velocities observed along PA=22$^\circ$ define a flat
rotation curve for this galaxy. If the constant velocities of order
29~km~s$^{-1}$ along the major axis at PA=22$^\circ$ are due to pure
circular rotation then we should find constant velocities of order
14,4,9,13~km~s$^{-1}$ along
PA=59$^\circ$,130$^\circ$,146$^\circ$,160$^\circ$ respectively.  We
don't have any evidence for these constant values in the inner region
except possibly for PA=160$^\circ$.  Moreover, we find no acceptable
combination of mass components (bulge, dark matter halo, stellar disk)
for fitting a completely flat curve in the innermost 0.5~kpc region of
M33 and its sudden rise around 0.5~kpc. In
addition, the disk mass to light ratio in the innermost regions is
constrained by the fit to the large scale RC and cannot be considered
arbitrarily small such as to leave a dominant shallow bulge component
out to 0.5~kpc.

In this sub-section we still want to examine another possibility
related to the axisymmetric potential, namely that the rotation curve
is traced by the CO and HI lines down to 0.3~kpc. We consider only the
two innermost points of our optical binned emission line set to
complement this further in.  This removes our previous assumption of a
flat rotation curve out to R$\sim 0.5$~kpc and that major axis
position angle is strictly 22$^\circ$.  The higher azimuthal averages
of CO velocities considered might in fact indicate that indeed the
rotational velocities are higher along other directions (consistent
with our data).  In Figure 10 $(a)$ the circled symbols show the inner
data used to define a rotation curve and which we now use fit a mass
model.  For $R>1.2$~kpc the rotation curve used is described by
Corbelli (2003) and extends out to 19~kpc.  We follow the approach
of Corbelli (2003) in a cold dark matter framework and use a dark
matter profile as described by Navarro et al (1997).  As mass
components we shall use an exponential stellar disk, a dark matter
halo, a bulge, a gaseous disk and the nuclear cluster.  We allow the
exponential disk to have all possible mass-to-light ratio values. For
the bulge we assume a rotational velocity of the following form:

\begin{equation}
V_{sph} = G M_{sph}  R^{0.5}/(R+s) 
\end{equation}

\noindent where the mass $M_{sph}$ and the scale radius $s$ are free
parameters.  We set the stellar disk scalelength equal to the K-band
scalelength of 5.6$'$ (Regan $\&$ Vogel 1994) and search for the bulge
and dark matter parameters which give the lowest $\chi^2$ compatible
with the data.  The lowest bulge mass and scalelength $s$ compatible
with the data are $9\times 10^7$~M$_\odot$ and $s=0.15$~kpc.
We show the central region of this mass model fit in Figure 10$(a)$.
The mass to light ratio of the disk and of the bulge are of order
unity even though uncertainties in the bulge luminosity leave this
ratio uncertain for the bulge itself. A larger bulge gives acceptable
fits and we cannot exclude it from dynamical arguments. The dark
matter halo parameters in this fit are similar to those found by
Corbelli (2003).  The gas rotation curve in this case is far from
flat and it is compatible with the rotation curve inferred from the
stellar absorption lines, provided that the last one is affected by
asymmetric drift.  The velocity dispersions required by the asymmetric
drift correction to stellar rotation in order to agree with gas
rotation are of order 20-30~km~s$^{-1}$.

Could the position angle of the major axis be effectively changing for
$0.2< R<0.6 $~kpc, as indicated by higher velocities observed in CO
along other directions? We cannot exclude that the direction of major
axis is changing with radius but this alone does not explain the
complex pattern we observe in the emission line velocities.  Despite
the good dynamical fit to the slow rising rotation curve of Figure
10$(a)$, compatible with the dynamical model of the extended disk
rotation, its projection along other directions does not agree with
the observations presented in this paper.  Orthogonal directions such
as PA=146$^\circ$ and PA=59$^\circ$ reach similar velocities on the
East side for example but much lower value on the West side.
Overcoming these discrepancies in the framework of simple models
discussed above implies taking into account additional peculiar
motions. M33 has its near side west of major axis. Evidence for this
is provided by dust lanes, which become more pronounced west of
the major axis in a blue image of M33 (e.g. Figure 1). This
orientation is also consistent with trailing spiral arms.  In this
geometry, a positive anomalous velocity on the near side can be
interpreted in terms of local radial infall of gas towards the
center. Evidence for local radial gas infall in our sampled areas
might come only from a region 1$'$ west on minor axis and along the East
side at PA=59$^\circ$.  These peculiar velocities might be connected
to shocks along dust lanes or to the beginning of the northern/southern 
spiral arm.  The possibility that these features are related to
shocks will be analyzed in more detail in a forthcoming paper where
emission line ratios are examined.  South-East of the nucleus, towards
the strong OB association, the gas shows strong positive velocities
which could be interpreted as a large nuclear outflow, possibly
intersecting also the innermost 1$'$ East on minor axis.  However,
since a similar anomaly is observed also in the stellar velocity
pattern, even though of smaller amplitude, we cannot give any definite
conclusion on the existence of such outflow.

In the rest of this Section we will discuss the possibility that the
anomalous velocities we observe are not related to gas inflow or
outflow patterns superposed on pure circular rotational velocities
in an axisymmetrical potential, but to large scale distortions of the
potential, such as given by a displaced bulge or by a weak bar.

\subsection{A displaced bulge}

The ordered motion detected in central regions of M33 and the high
star formation rate (Thilker et al. 2005; Magrini, Corbelli, $\&$
Galli 2007) point out that if any bulge is in place this is more
likely a pseudo-bulge, intermediate between a spheroid and a disk.

Looking in detail at the gas and stellar radial velocities along
PA=130$^\circ$,146$^\circ$,160$^\circ$ we notice that between -1$'$
and 0.5$'$ velocities are symmetric around a point located at
$R\simeq -0.25'$ where velocities are of order
-170~km~s$^{-1}$. Abrupt changes in the velocity are noticeable
outside the -1$'$ -0.5$'$ region. These asymmetries around a displaced
center with respect to nuclear cluster are possibly in line with a
suggestion by Minniti et al. (1993) about a bulge that is offset from
the nuclear cluster. They reached this conclusion from looking at the
V and H band images of the center of M33.  However the presence of
dust lanes in this region and of the bright OB association in the
south-east did not allow any definitive conclusion to be drawn. The
displacement in the kinematic symmetry point we find is of order
15$''$ to the South-East, compatible with the suggestion of Minniti et
al. (1993). The excess light they found in the innermost 0.24~kpc, is
compatible again with what is required by the mass modeling of the
large scale rotation curve.  The excess of light in the innermost
0.5~kpc region detected by azimuthal averages of near infrared J,H,K
emission by Regan $\&$ Vogel (1994) points to a larger spheroidal
component (8$'$ effective radius).  This conclusion however relies on
the assumption that the exponential disk scalelength does not change
going from the extended disk to the innermost region and that an
$r^{1/4}$ light distribution is effectively in place for the
spheroid. In reality, departures from $r^{1/4}$ law are evident both
in the blue and in the near infrared bands (Bothun 1992; Regan $\&$
Vogel 1994).  A possible displacement of the bulge however does not
explain the peculiar symmetric steep rise and flatness of the gas
velocity observed along PA=22$^\circ$ centered on the stellar nuclear
cluster.

\subsection{An oval bar}

As stated in the Introduction, photometric evidence for a bar has been
given by near-IR photometry but also by the Fourier transform
technique on galaxy images in the blue band (Elmegreen, Elmegreen $\&$
Montenegro 1992). We shall discuss here whether any anomalous velocity
with respect to the axisymmetric potential can be interpreted in terms
of a triaxial potential, and if this triaxial figure is rotating, as
is the case for a bar. Near-IR images point to a bar which we
initially estimated to be roughly aligned along PA=96$^\circ$. 
This position angle estimate is close to
the minor axis and that makes the bar more difficult to see because
projection effects make it appear rounder. To investigate the light
distribution we downloaded the 2MASS K-band image, rotated it by 
22$^\circ$ clockwise to align the major axis with y-axis, and stretched
the x-pixel size by 1.62 to attempt to deproject the image to what it
might look face-on if the components are all in a thin disk.  The
central portions of the rotated and of the deprojected images are
shown in Figure~11. The stretched image needs to be interpreted with
caution since e.g. a spherical light distribution would be stretched
out into an oval shape. Nevertheless, a central broad oval
concentration of light appears to be visible in the non-stretched
image too.

We measured the ellipticities of the isophotes on a non-stretched
median filtered K-band image. The ellipticity is low averaging 
about 0.15 from 50 to 150~arcsec distance from the center, and the PA
of the isophotes is $\simeq 80^\circ$ in the non-rotated frame,
somewhat different from our previous estimate 
(96$^\circ$ by inspecting stretched images). Inside
50~arcsec but still outside the central star cluster the ellipticity
is higher, about 0.3. Outside 150~arcsec visible spiral arms
converging onto the center limit the validity of the isophote fitting
procedure.  The deprojected bar radial extent can then estimated to be
at least 2.5$'$ or 0.6~kpc, and it can possibly be even larger.  Regan
\& Vogel (1994) estimated a deprojected bar length of about 0.8~kpc
while Elmegreen et al. (1992) claim a longer bar, extending out to
about 1.5~kpc.  It is not yet clear what limits the extent of weak
bars.  According to Combes $\&$ Elmegreen (1993) in late-type galaxies
the bar extent is comparable to the radial scalelength of the stellar
disk; the K-band radial scalelength of the stellar disk of M33 is
about 1.4~kpc\citep{reg94} so this theory predicts a bar length in
close agreement with the estimate of Elmegreen et al. (1992).  The bar
in M33 is weak according to most classification schemes.  For example
following Abraham et al. (1999) and using the 2-MASS image of Figure
11 we computed an intrinsic bar axial ratio of 0.57, and a bar
strength parameter $f_{bar}=0.2$. These values place M33 at the
boundary between barred and unbarred systems and marginally classify
M33 as a barred galaxy. Measuring and interpreting the inner
kinematics is therefore especially important for this galaxy in order
to find kinematic signatures of the oval bar.

The region along the minor
axis where we see anomalous velocities has an extent of about 
1.2$'$ on the sky  i.e. of about 0.5~kpc in radius. This is close to
the smallest effective bar length estimate. The positive velocities of
the gas East of the star cluster on the minor axis, and the negative
velocities on the opposite side, at about 8$''$ away from the stellar
cluster, can be interpreted in term of streaming motion of elongated
orbits parallel to the bar. It is likely in fact that streaming
motions become more pronounced closer to the center, where light
contours are more elongated.  The remarkably constant velocities for
the emission line gas observed on both sides of the nucleus at
PA=22$^\circ$ are also consistent with streaming motions along a bar:
For a bar oriented close to the minor axis as is the case for M33,
streaming motions along the bar add to the circular velocities on the
major axis to produce a sharp discrete jump in velocity across the
nucleus. This is seen  clearly in the galaxy NGC6300 which
has a bar close to the minor axis and inclination similar to M33 (Buta
etal. 2001, their Figure 21). Major and minor axis velocity
peculiarities in the inner $\pm 1'$ are then consistent with having
streaming motions along a bar.

Let us now assume that the central flat plateau of the velocity
curve along PA=22$^\circ$ is mostly due to streaming motion and that
the average rotational velocities are lower, somewhere in between the
stellar velocities and the CO velocities. For fitting a mass model we
take the average velocities traced by optical emission lines 
along the northern and southern side of the major axis between 0.2 and 
0.8~kpc. We consider the
rotation curve of Corbelli (2003) at larger radii.  In Figure 10$(b)$
we show with circled symbols the data points in the inner region which
we shall use to fit a mass model.  Since we don't require rotational
velocities to be high at $R<0.2$~kpc, we can achieve acceptable fits
without putting any relevant mass in the bulge. Best fitting
parameters for the stellar disk and for the dark matter halo are
similar to those of the model shown in Fig. 10$(a)$. The
additional non axisymmetric mass component in the bar is what causes
the streaming motion i.e. the larger velocities observed in the gas
(at $R<0.2$~kpc along the major axis and in the azimuthal
averages traced by the molecular component out to 0.8~kpc) 
with respect to the pure circular rotation (continuous line in Figure 
10$(b)$). The difference
between the gas and the stellar rotation curve in this case is
compatible with the asymmetric drift hypothesis even though in the
presence of a bar, stellar and gas orbits might be different.

Using this rotation curve and the epicyclic approximation we can then
locate the inner/outer Lindblad resonances, (hereafter ILR/OLR) and the
corotation radius, if a bar pattern speed can be inferred. The
positive radial gradient of the rotation curve shown in Figure 10$(b)$
implies that there is no ILR or that this is close to the cluster
location, at very small radii. In Figure~12 we show the angular
frequency $\Omega$ corresponding to the rotation curve of Figure
10$(b)$.  After computing the epicyclic frequency $\kappa$ we draw the
$\Omega \pm \kappa/2$ curves needed to locate the inner and outer
Lindblad resonances.  If the bar in M33 ends at its own corotation
(0.6-1.5~kpc) then the bar pattern speed is 10$\pm 2$~km~s$^{-1}$
(shaded region in Figure 12). The intersection between the $\Omega
-(+) \kappa/2$ curve and the bar pattern speed defines the ILR (OLR).
The requirement of a bar ending at the IRL would imply instead a
very low bar pattern speed, of order 1-2~km~s$^{-1}$.  The absence of
an ILR (possibly occurring at very small radii), or a low bar pattern
speed, confirm the weakness of the M33 bar. The likely situation is
that the bar of M33 is at its dawn, and still in the process of
aligning the orbits in a preferential direction.  Our finding is
linked to the argument of Binney $\&$ Tremaine (1987) about the growth
of weak bars: namely, that when a weak bar instability arises in a
disk, most of the orbits are aligned parallel to the bar, streaming
motion are large in the direction perpendicular to the bar and there
is no ILR.

Evidence for weak bars or triaxial structures has been
found for two other late-type spirals recently: NGC 628 (Fathi et
al. 2007) and NGC 2976 (Spekkens \& Sellwood 2007). They show
that late-type spirals may not be the simple dynamical systems we once
might have thought them to be. In the case of M33, streaming motions
are difficult to quantify given the small amplitude of the observed
velocities and the small size of the region where these motions take
place.  Even though a weak bar can likely drive non-circular motion
and cause the non zero velocities along minor axis and the flat
velocity profile along the major axis of M33, our impression is that
it cannot account for all peculiar velocities observed in other
directions.  Any more detailed model for a non axisymmetric flow
pattern requires a full 2-D velocity map of the central region.

\section{Summary and discussion}

In this paper we have presented and analyzed optical observations of
gas and stellar radial velocities in the innermost 0.5~kpc region of
M33. We summarize below the main results and their
implication for the mass distribution in the central region of this
nearby galaxy.

\begin{itemize}

\item[({\em i}\/)]The radial velocity of the bright nuclear
cluster places it close to the systemic velocity of this galaxy.
We see only a small offset in the dynamical center as determined
from the required symmetry in emission line radial velocities 
across the central steep velocity gradient observed along PA=22$^\circ$
(major axis of the large scale disk). The lack of a dominant
gravitational entity in the center of this
galaxy implies that the exact dynamical center cannot be yet defined.

\item[({\em ii}\/)]Gas radial velocities indicate a relatively flat 
and symmetric curve along PA=22$^\circ$ between 4$''$
and 2$'$ radial distances from the nuclear cluster. Some
local displacements around the mean value of 22~km~s$^{-1}$ are visible.
No other direction, between all those observed, presents evidence 
for a very steep central gradient 
and for a flat and symmetric velocity distribution such as observed 
along PA=22$^\circ$ (about a 45~km~s$^{-1}$ jump over a few pc).
For each slit crossing the nuclear cluster the maximum observed radial 
velocity however is of order 
20-30~km~s$^{-1}$ (as along PA=22$^\circ$) but it can occur as far as 
1.5$'$ from the nuclear cluster. Particular relevant are the non
zero velocities observed for the gas along the minor axis within
projected radial distances of  $\pm 1'$. Given the large scale
orientation of the M33 disk these 
cannot be explained in terms of a gas inflow pattern.

\item[({\em iii}\/)]The stellar velocities show smaller radial
gradients than the gas, even though the overall pattern
along each direction agrees with that of the gas. We have discussed
the differences between the gas and stellar velocities in terms
of asymmetric drift correction. We estimate that the asymmetric drift
correction cannot explain the difference between a flat gas rotation
curve and a rising stellar rotation curve for an axisymmetric
potential. Moreover we have not found a mass distribution compatible
with a flat rotation curve inside 0.5~kpc region and with the large scale
rotation curve.

\item[({\em vi}\/)]We have discussed the likely possibility that the
flat gas velocity curve observed along PA=22$^\circ$ and the non
zero velocities observed  in the orthogonal direction are connected
with streaming motion due to a weak bar. This bar has been postulated
previously from photometric measures but here we have
presented the first dynamical evidence consistent with this. 
With M33 included, it now seems all major Local
Group galaxies (including the Large Magellanic Cloud) are barred!


\end{itemize}

The presence of a weak bar and the uncertainties in the
amplitude of the streaming motion close to the center of this galaxy
limit our ability to constrain the density profile of the dark matter
halo in the center and any related cosmological issues.
The other limit comes from what Figure 10 shows, namely that
the likely baryonic mass components in the central region of M33 are
never so small as to leave dark matter as the main ingredient of the
gravitational potential. The presence of relevant baryonic components
with their related uncertainties limits the ability to constraint the
shape of dark matter density distribution.
If a galaxy like M33 were 5-10 times further away,
the non-circular motions we have found might have gone unnoticed. In
this sense, M33 provides a sobering reminder that even a nearby
disk-dominated system still does not provide a unique answer to this
cosmological issue.

\acknowledgments We thank the referee for her/his useful comments
to the original manuscript and Josie Fenton for her work on an initial 
data set taken with the first version of the DIS spectrograph before its
upgrade to new CCD detectors. This paper is
based on observations obtained with the Apache Point Observatory 3.5-m
telescope which is owned and operated by the Astrophysical Research
Consortium. We appreciate the support from the APO
observatory staff and the support, both technical and financial, that
has led to upgrades to DIS that have been very beneficial for this
project. This publication makes use of data products from the 2MASS, 
which is a joint project of the University of Massachusetts and 
IPAC/California Institute of Technology funded by the National
Science Foundation and the National Aeronautics and Space Administration.

\newpage

\begin{figure}
\includegraphics[width=85mm]{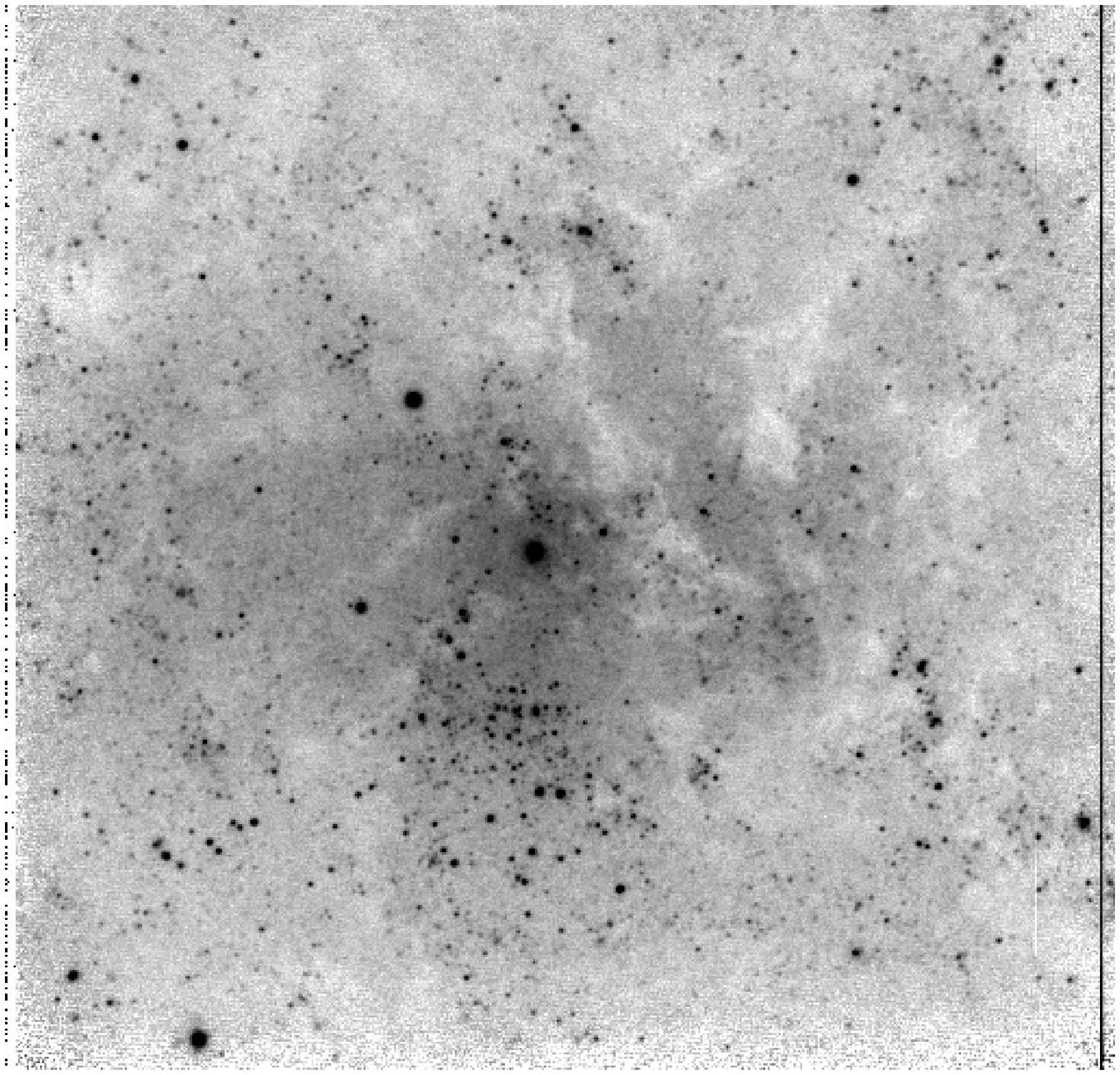}
\includegraphics[width=85mm]{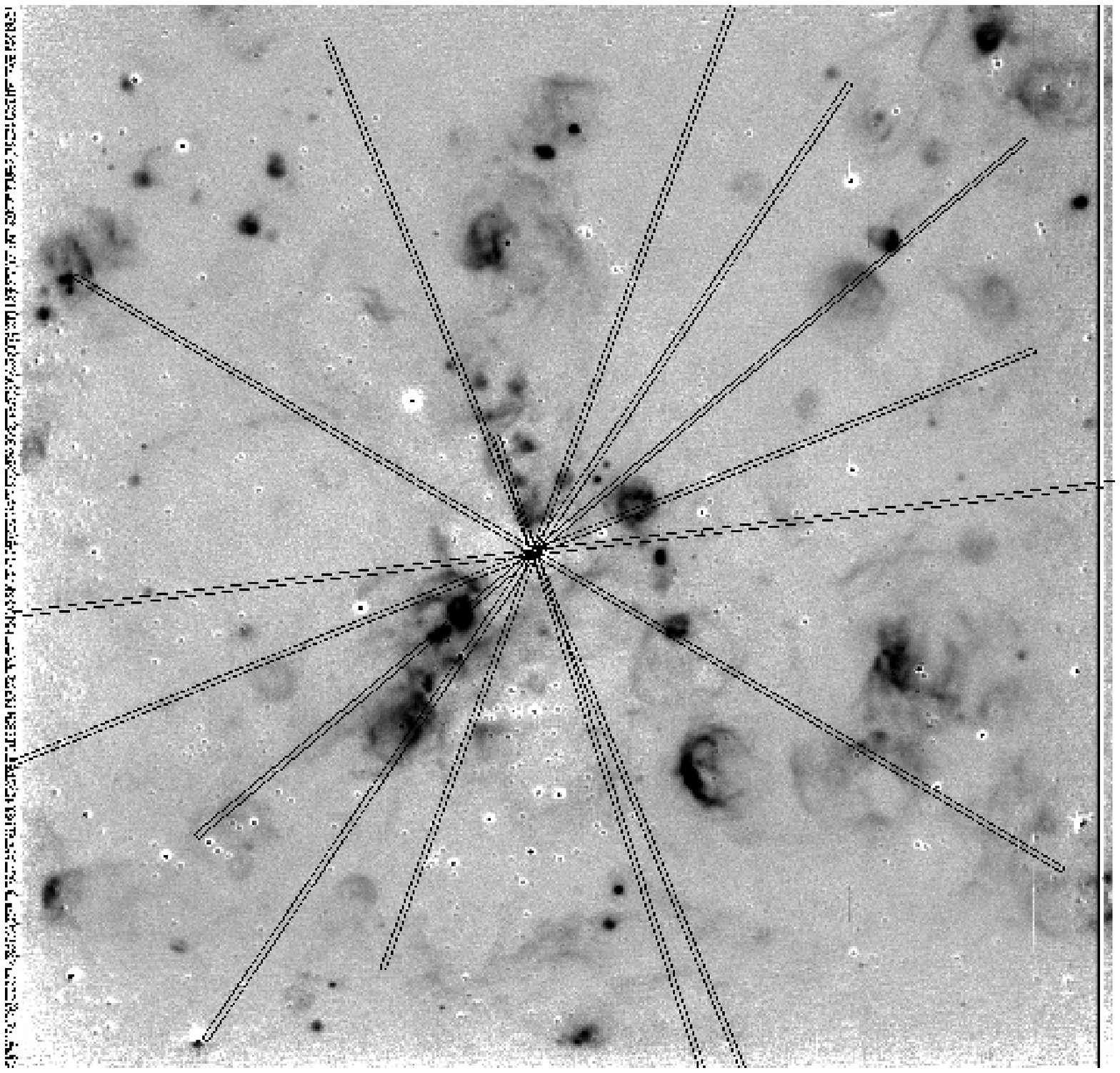}
\caption{\label{fig1}$Left$ Blue image of the central region 
of M33 obtained with the 
ARC 3.5m telescope through a 100\AA\ wide filter centered at 
4750\AA. M33's disk has a position angle of 22$^\circ$. The area to 
the right of the major axis is the near side which may explain why dust 
clouds are best seen there. $Right$ H$\alpha$ image of the same
region. The continuum emission has been subtracted from this
image. The various slit positions at which spectra were obtained are
shown. They are, anti-clockwise from top, PA=18$^\circ$,22$^\circ$,
59$^\circ$,96$^\circ$,112$^\circ$,130$^\circ$,146$^\circ$, and 160$^\circ$. 
In addition to the bright HII regions, there is diffuse ionized
gas present across the region, enabling measurements of emission line
velocities throughout each slit.
North is up and East is left in both images. The frames measure
about 280$''$ on the side, where 1$''$ corresponds to 4~pc at 840~kpc distance.}
\end{figure}

\begin{figure} 
\includegraphics[width=124mm]{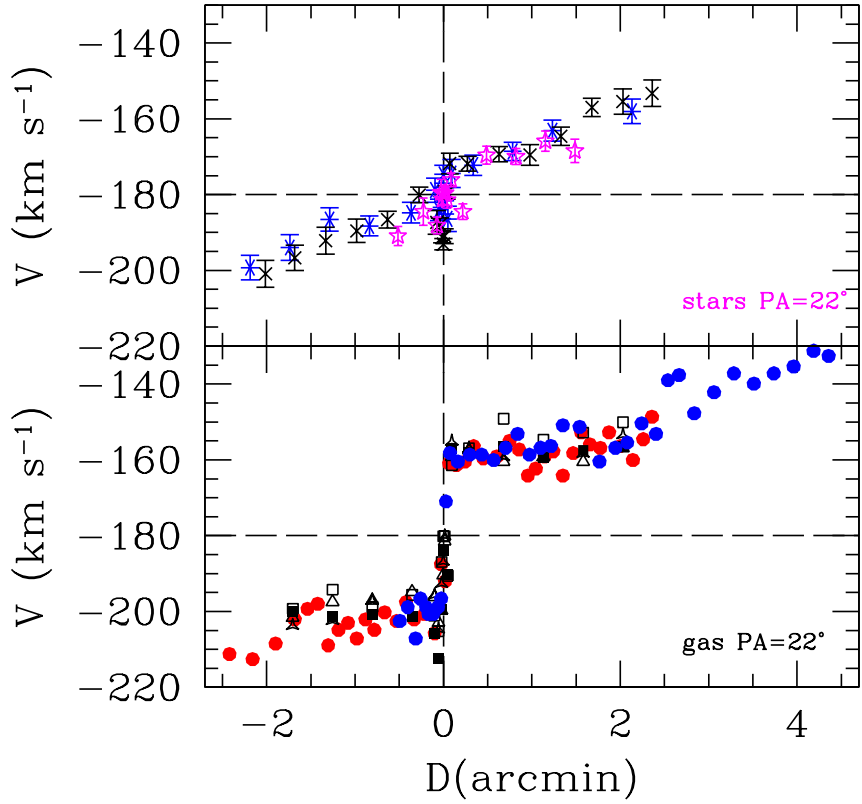} 
\caption{\label{fig2}$Bottom~panel$. Radial
velocities of the ionized gas in km s$^{-1}$ along PA=22$^\circ$, the nominal
major axis, as a function of projected distance (D) from the center (nuclear
star cluster) in arcmin ($1'$=240~pc). 
Filled circles refer to simultaneous fits of the 4
red emission lines, filled squares to H$\alpha$ measurements, open
squares to [NII]6584, open triangles to [SII]6716, and inverted "Y" to
[SII]6731. Note the flat velocity profile and sudden velocity jump
across the central region which occurs over just a few arcsec (see
also Fig.7 and Fig.8). The vertical dashed line marks the position of the star
cluster in this and subsequent panels.  $Top~panel$. Stellar radial
velocities along the same PA derived from the Ca triplet (open star symbols) 
and blue spectra (cross and asterix symbols, each relative to one observing
run). Stellar velocities show a smoother increase with distance from the center 
than ionized gas velocities.}  
\end{figure}
 
\begin{figure}
\includegraphics[width=124mm]{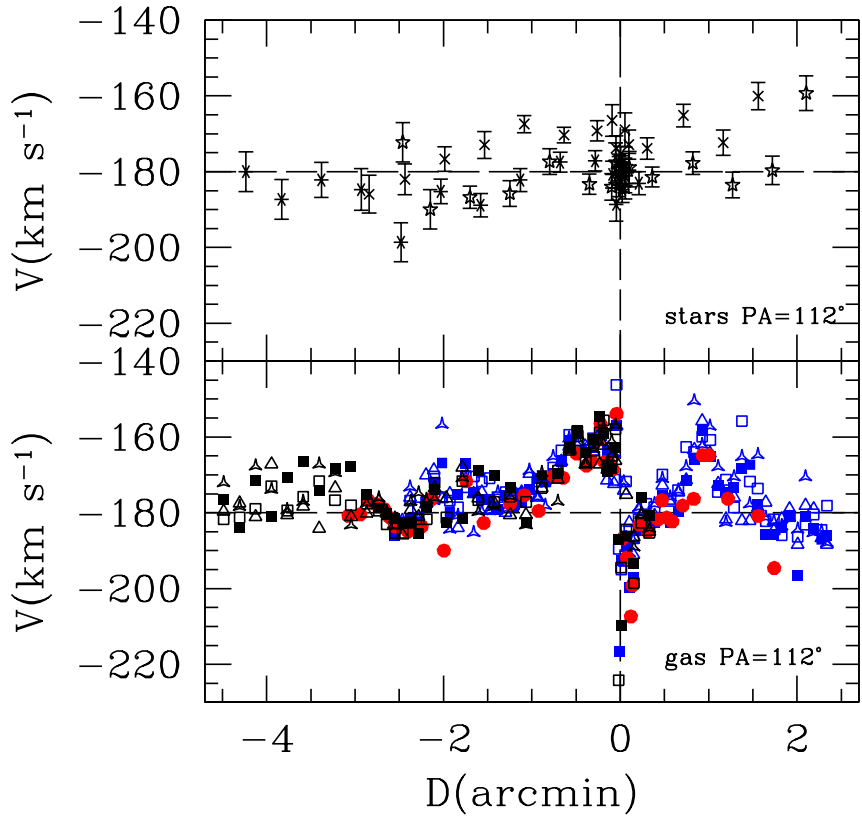}
\caption{\label{fig3} $Bottom~panel$. Radial velocities  
of the ionized gas along PA=112$^\circ$, the nominal minor axis, 
as a function of projected distance from the center. 
Symbols as defined in Figure 2. Notice the strong non-zero velocities 
out to a radius of about 1$'$ on each side.
$Top~panel$. Stellar velocities on the minor axis. Different symbols refer to 
velocities derived from correlation technique on the Mg side of blue spectra 
obtained during different observing runs. Average 
stellar velocities seem higher than V$_{sys}$ = $-180$ km~s$^{-1}$ in
those regions where gas velocities are also higher. 
}
\end{figure}

\begin{figure}
\includegraphics[width=124mm]{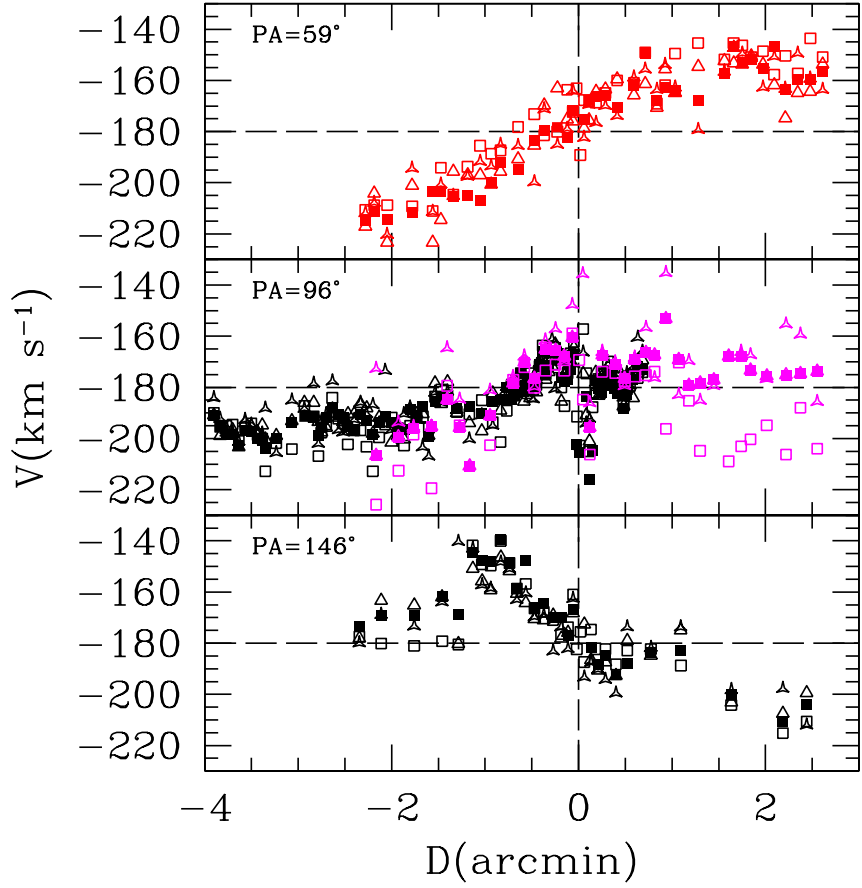}
\caption{\label{fig4} From $top$ to $bottom$: Radial velocities of 
the ionized gas along PA= 59$^\circ$, 96$^\circ$, and 146$^\circ$. 
Symbols as defined in Figure 2.  The regions with peculiar velocities are 
again confined to $\pm$ 1-1.5$'$ from the center.}
\end{figure}

\begin{figure}
\includegraphics[width=124mm]{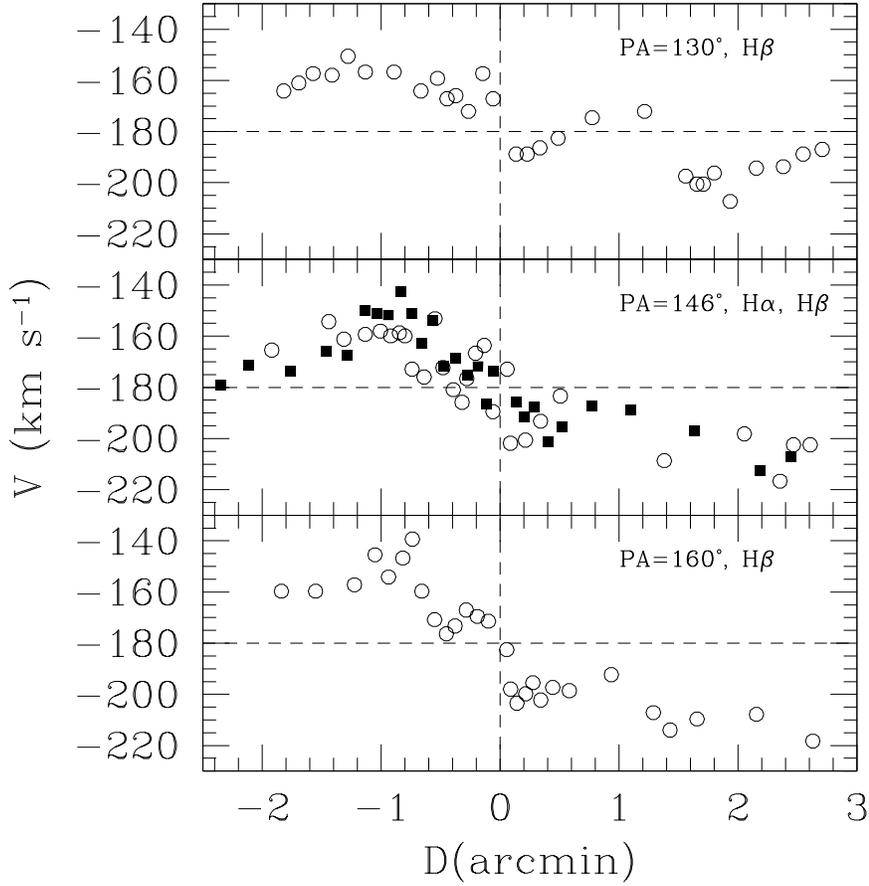}
\caption{\label{fig5}
Radial velocity measurements for the emission line gas at
PA=130$^\circ$, 146$^\circ$, and 160$^\circ$. For PA=146$^\circ$, 
the 2005 H$\alpha$ and 2006 H$\beta$ (open circles) measurements are
shown. Zero points for the H$\beta$ measurements have been
determined through comparison with  PA=146$^\circ$ H$\alpha$
velocities. 
}
\end{figure}

\begin{figure}
\includegraphics[width=124mm]{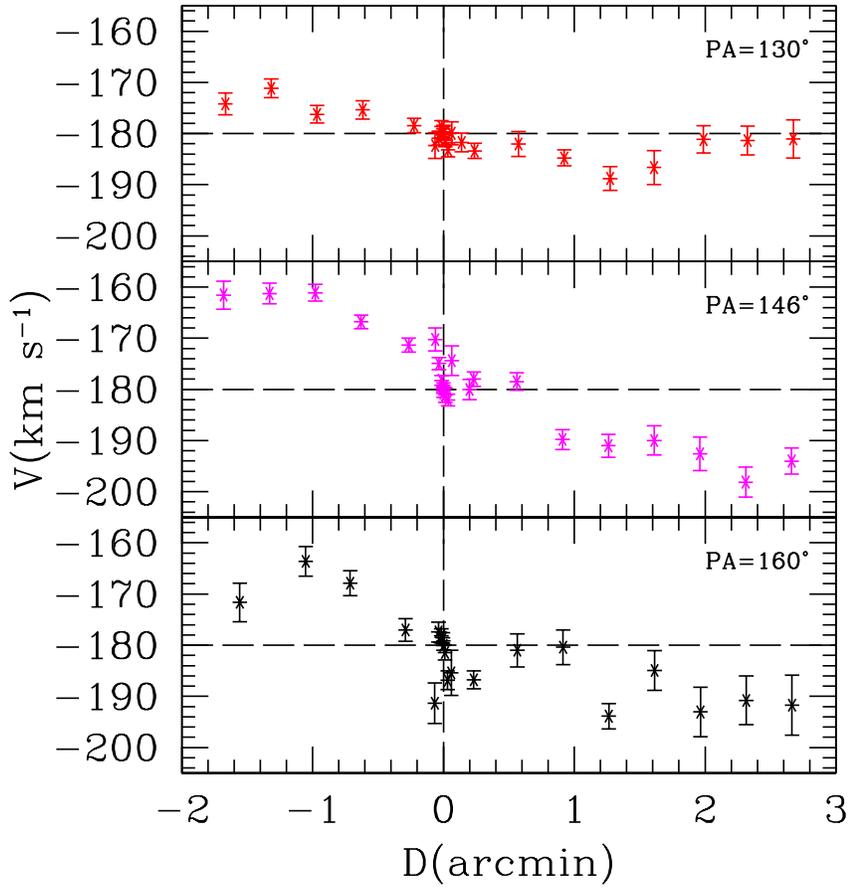}
\caption{\label{fig6}From $top$ to $bottom$: Radial velocities of 
stars along PA= 130$^\circ$, 146$^\circ$, and 160$^\circ$
from cross-correlation with template star spectrum on the Mg side
wavelength range. The nuclear cluster is centered at -180~km~s$^{-1}$.
}
\end{figure}

\begin{figure}
\includegraphics[width=144mm]{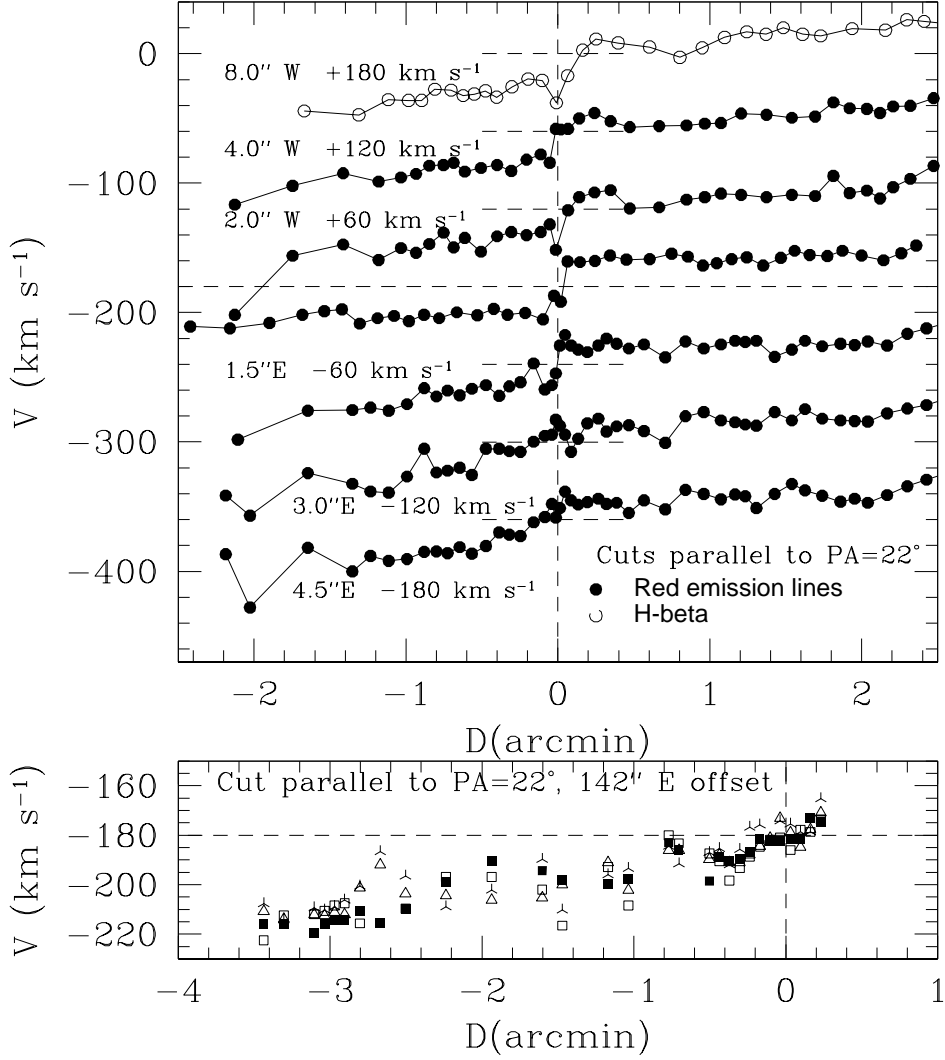}
\caption{\label{fig7}
$Top~panel$: Emission line velocities observed along the major axis
(PA=22$^\circ$) and along
slit positions parallel to it. NE is to the left and SW
to the right.  Open circles refer to H$\beta$ lines, filled circles
to averages over the four red lines.
The perpendicular distance from the major axis axis for
each cut is listed, as well as the velocity offsets added to separate
the curves. The dashed lines indicate the systemic velocity and location 
of the center. 
$Bottom$: Radial velocities measured along a slit position
parallel to the major axis but offset 142$''$ in the SE direction along
the minor axis. The intersection with the minor axis is shown. The
points around -3$'$ are centered on a bright HII region in the disk. }
\end{figure}

\begin{figure} 
\includegraphics[width=134mm]{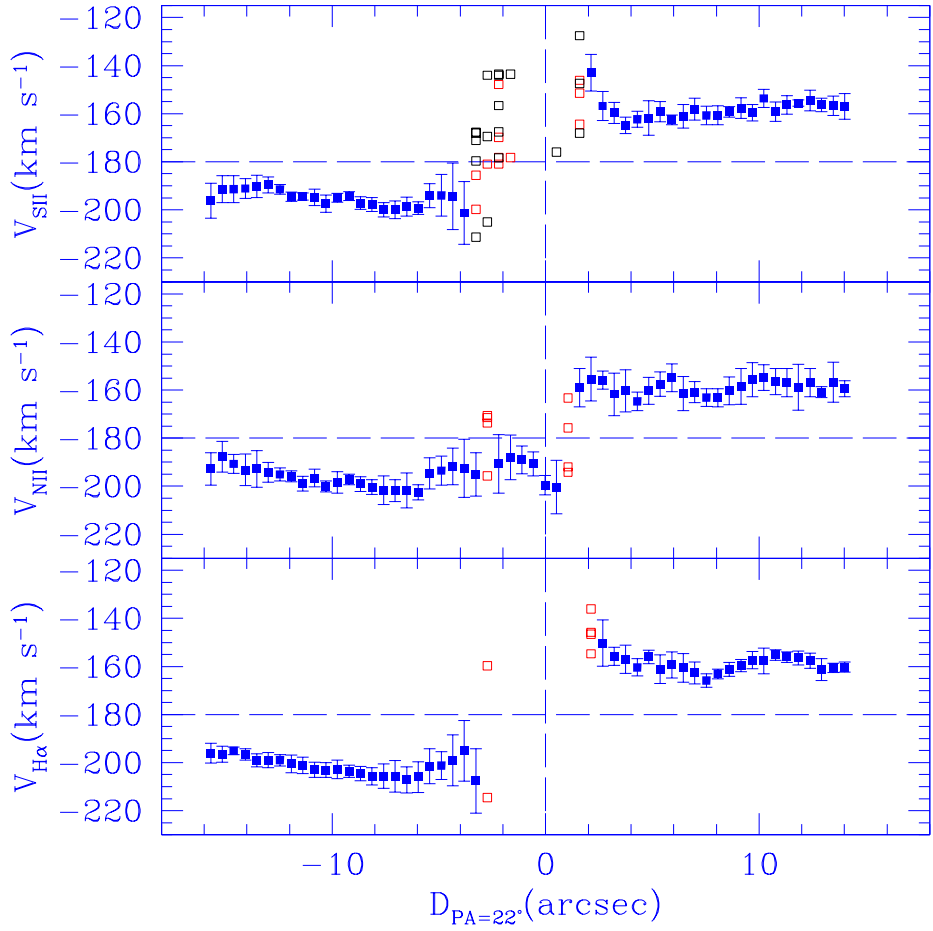} 
\caption{\label{fig8}Gas radial velocities along
the nominal major axis, PA=22$^\circ$, in the innermost 
region of M33. Lines have been measured
in single frames and pixel by pixel. The spatial sampling is 0.54$''$
but the typical seeing was about 1.5$''$ so neighboring points are not
independent.  Filled squares correspond to average values of the
velocity from measurements in different frames. Averages are not
computed when line velocities from different frames are inconsistent
i.e. when their dispersion around the mean (errorbars) is larger than
25~km~s$^{-1}$. In this case and when lines are hardly measurable due
to blending with stellar absorption or to intrinsic weakness, single
frame measurements are plotted (open squares).  }  
\end{figure}

\begin{figure}
\includegraphics[width=134mm]{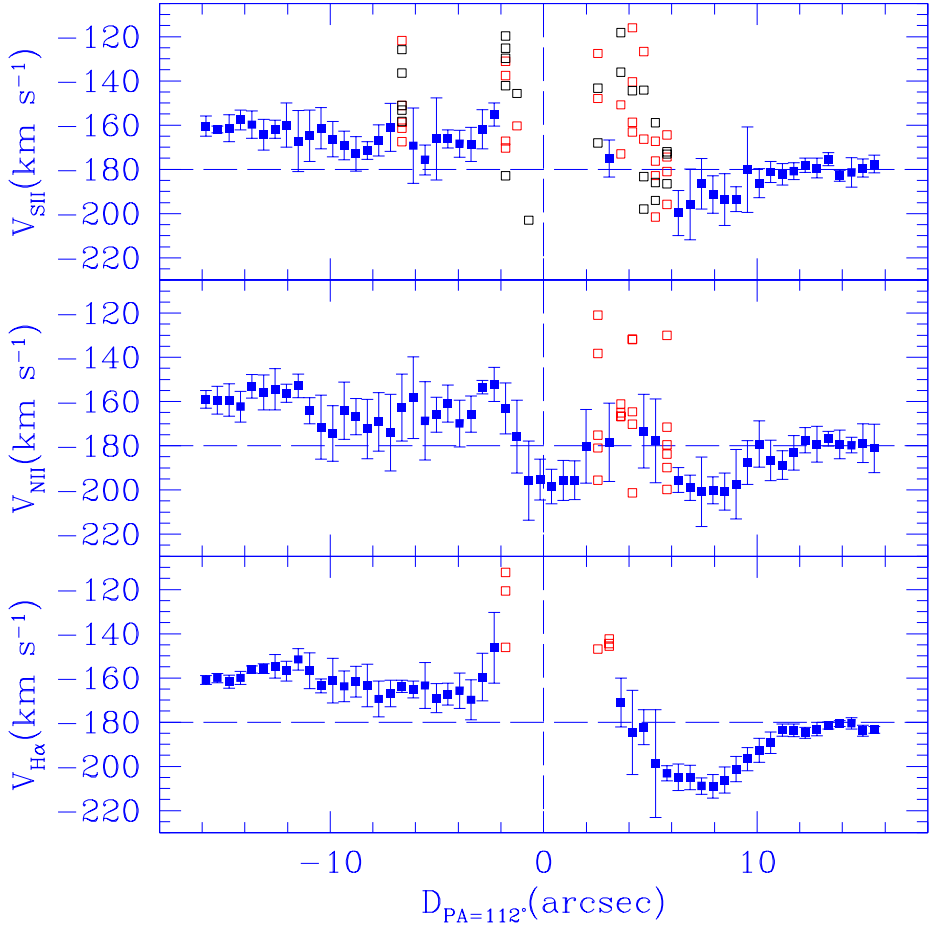}
\caption{\label{fig9}Same as Figure 8 but for the minor axis, PA=112$^\circ$.}
\end{figure}

\begin{figure}
\includegraphics[width=124mm]{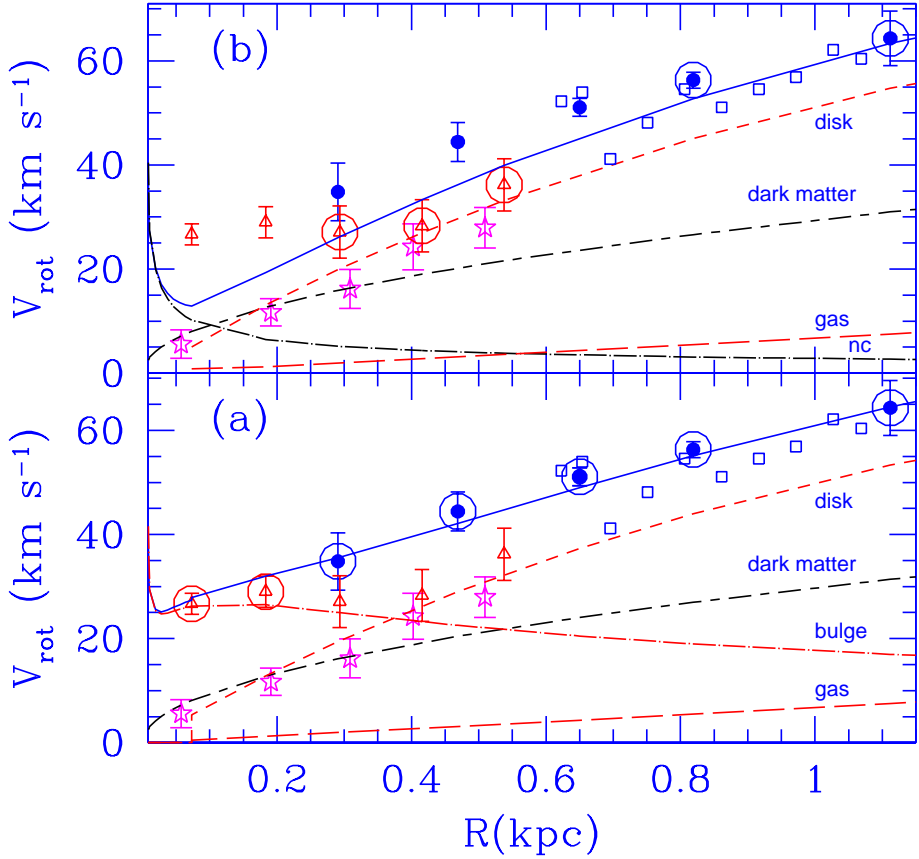}
\caption{\label{fig10} Rotational velocities assuming 
PA=22$^\circ$ for the major axis direction and $i=52^\circ$.
Triangles are gas average velocities traced by optical emission lines in the   
red and in the blue frame along PA=22$^\circ$. 
Errorbars indicate the uncertainties
in the mean in each bin. Open stars are stellar rotational
velocities along the same direction, measured using the MgII side of the blue 
frame only. Filled circles show the CO J=1-0 rotation curve from azimuthal averages 
of data points within $\pm45^\circ$ of major axis (Corbelli 2003). 
Open squares are 
from emission lines in one red frame along the south-west side of major axis. 
The continuous line shows the resulting best fit to the rotation curve defined 
by the circled symbols in this inner region and by data shown 
in Corbelli (2003) at larger radii. Dashed lines mark the contribution of the 
various mass components to the rotation curve assuming pure circular motion:
stellar and gaseous disk, bulge or nuclear cluster(nc), dark matter halo. 
In $(a)$ the fitted rotation curve is traced by optical emission lines 
inside 0.2~kpc and by the CO and 21-cm lines at larger radii.
In $(b)$ the fitted rotation curve is traced by averaged optical emission 
lines from 0.2 to 0.8~kpc and by CO and 21-cm lines further out. 
Inside 0.2~kpc optical emission lines are assumed to trace
the shear or non circular component due to a possible bar (see Section 4.3).
In both cases, $(a)$ and $(b)$, the dynamical model  
for pure circular motion predicts  rotational velocities which
are not constant but rise continuously outside 
the nuclear cluster region.}
\end{figure}

\begin{figure}
\includegraphics[width=85mm]{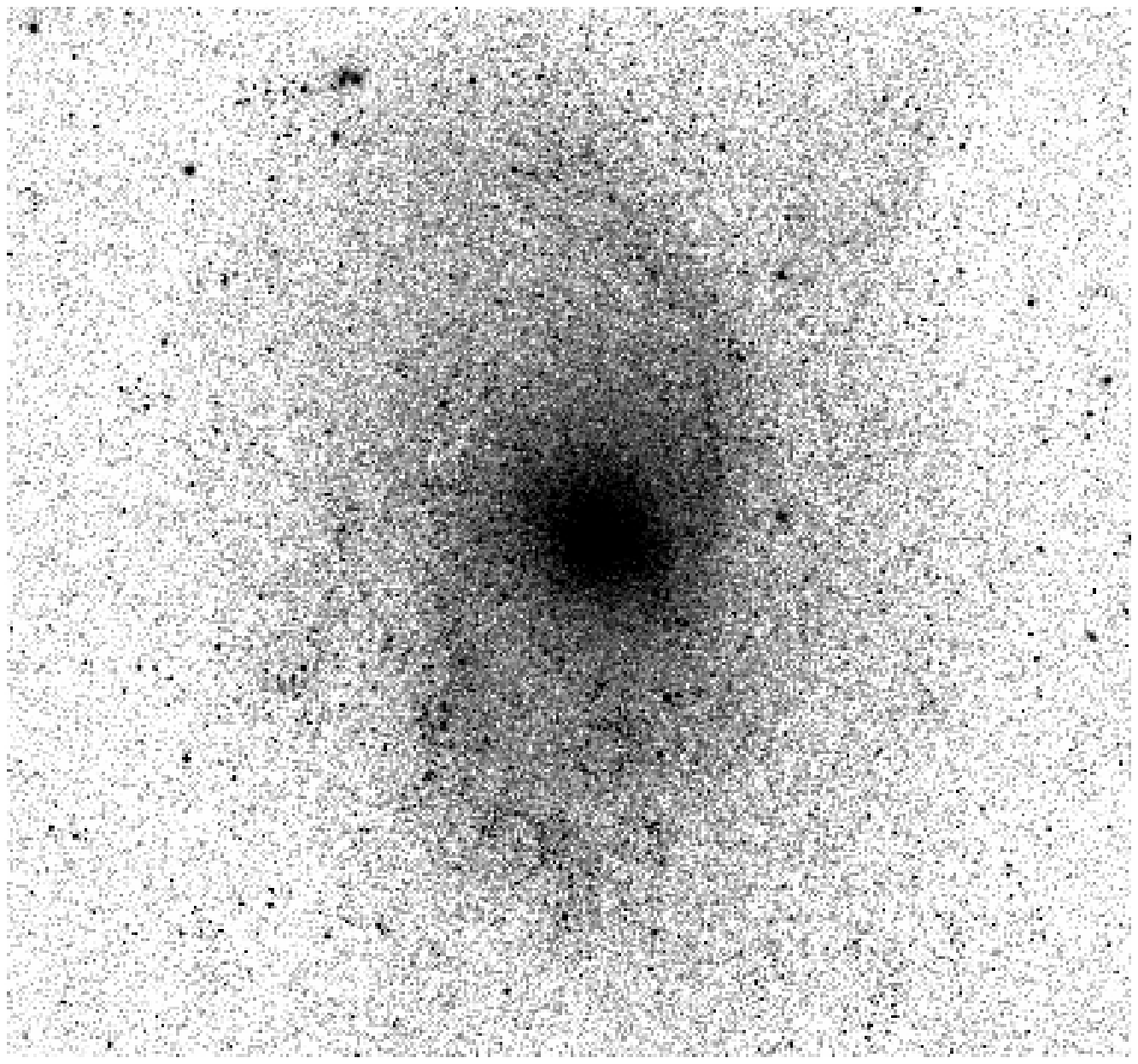}
\includegraphics[width=85mm]{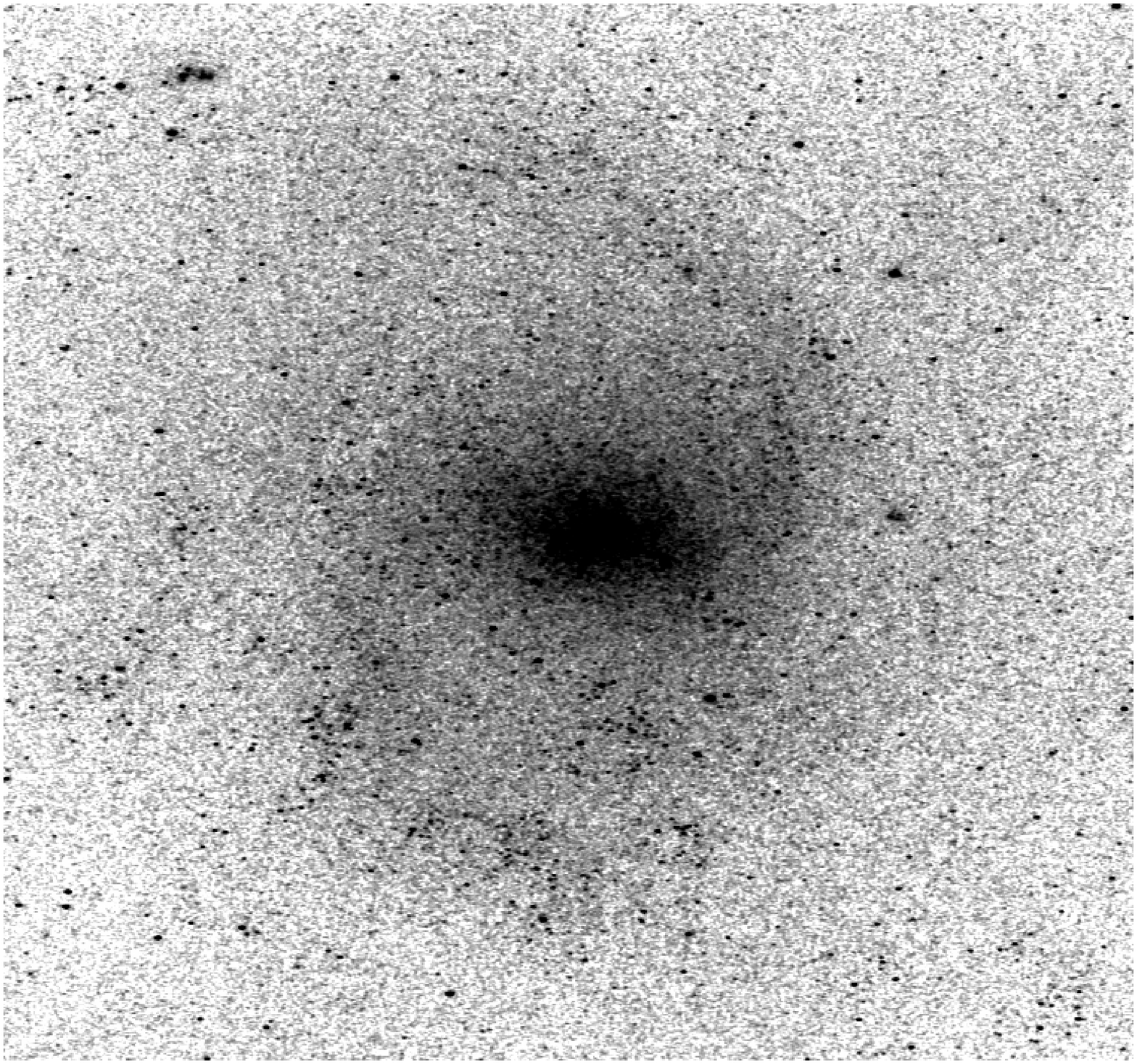}
\caption{\label{fig11} The left panel shows a 25'(horizontal) by 23.5' 
(vertical) region of the 2MASS K-band image of M33 rotated by -22$^\circ$ 
to align the disk's major axis with the vertical axis. 
The grey-scale stretch is chosen to emphasize the oval distortion 
of the light in the inner 3' diameter region. 
The right panel shows 
the same image deprojected to face-on by stretching the X-axis
by 1/cos(i=52$^\circ$). The deprojection is only representative of the face-on
light distribution if all components are intrinsically disk-like.
}
\end{figure}

\begin{figure}
\includegraphics[width=124mm]{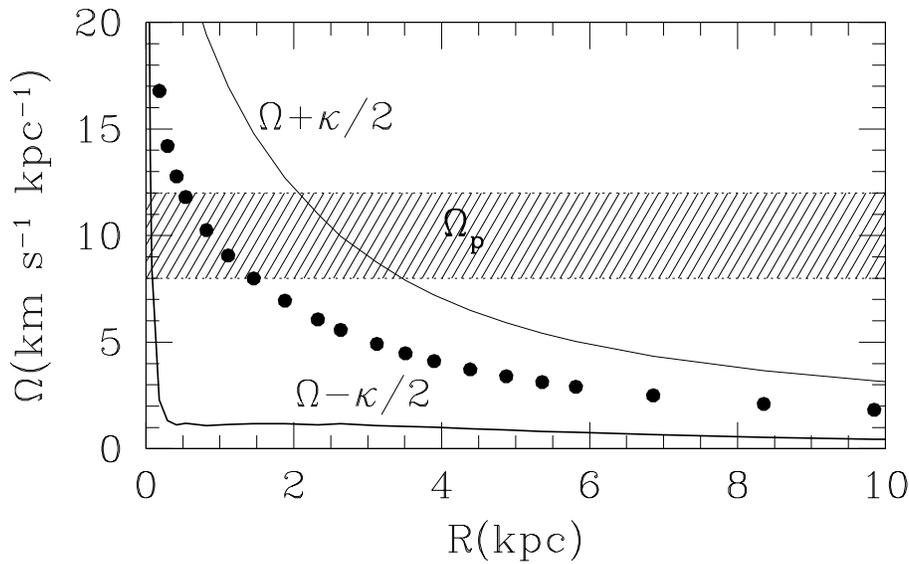}
\caption{\label{fig12} The angular frequency, $\Omega$, derived  from
the rotation curve of Figure 10$(b)$ (filled circles) and the
$\Omega \pm \kappa/2$ curves. The intersection between the latter curves
and $\Omega_p$ defines the location of the inner and
outer Lindblad resonances. The shaded region indicates the possible 
values of the bar pattern speed $\Omega_p$ if the bar ends at its 
own corotation.
}
\end{figure}

\end{document}